\documentclass[prb,showpacs,twocolumn,preprintnumbers,amsmath,amssymb]{revtex4}
\usepackage{epsfig}
\usepackage{graphicx}
\usepackage{dcolumn}
\usepackage{bm}
\begin{document}

\title{Scattering of electrons in graphene by clusters of impurities}
\author{M. I. Katsnelson$^1$}
\affiliation{$^1$ Institute for Molecules and Materials, Radboud
University Nijmegen, Heijendaalseweg 135, 6525 AJ, Nijmegen, The
Netherlands }
\author{F. Guinea$^2$}
\affiliation{$^2$Instituto de Ciencia de Materiales de Madrid
(CSIC), Sor Juana In\'es de la Cruz 3, Madrid 28049, Spain}
\author{A. K. Geim$^3$}
\affiliation{$^3$ Manchester Centre for Mesoscience and
Nanotechnology, University of Manchester, M13 9PL, Manchester, UK}

\begin{abstract}
It is shown that formation of clusters of charged impurities on
graphene can suppress their contribution to the resistivity by a
factor of the order of the number of impurities per cluster. The
dependence of conductivity on carrier concentration remains linear. In the regime
where the cluster size is large in comparison to the Fermi
wavelength, the scattering cross section shows sharp resonances as a
function of incident angle and electron wavevector. In this regime,
due to dominant contribution of scattering by small angles, the
transport cross section can be much smaller than the total one,
which may be checked experimentally by comparison of the Dingle
temperature to the electron mean free path.
\end{abstract}

\maketitle \section{Introduction.} Graphene currently attracts
intense attention as a novel, strictly two-dimensional system with
unique electronic properties that are interesting with respect to
both basic physics and potential applications (for review, see
Refs.\onlinecite{reviewGK,reviewktsn,NGPNG09}). It was shown already
in the early reports on graphene \cite{kostya1} that charge carriers
in this material exhibited a remarkably high mobility $\mu$ such
that submicron mean free paths were routinely achievable, and
an order of magnitude higher $\mu$ were observed for suspended
graphene samples \cite{bolotin2008,DSBA08}). Away from the
neutrality point, the conductivity of graphene is weakly temperature
dependent and approximately proportional to the carrier
concentration $n$ \cite{kostya2,kim}. Despite extensive experimental
and theoretical efforts, there is still no consensus about the
scattering mechanism limiting $\mu$ in graphene on a substrate.
Charged impurities are probably the simplest and thus the most
natural candidate \cite{NM06,Ando06,AHGS07}, and this conjecture is
in agreement with the experiments in which potassium atoms were
deposited on graphene at cryogenic temperatures \cite{CJAFWI08}.
However, room-temperature experiments with gaseous adsorbates such
as NO$_2$ have showed only a weak dependence of $\mu$ on charged
impurity concentration\cite{Schedin07}. The latter observation
agrees with several reports of only modest changes observed in $\mu$
after thermal annealing of spuriously doped samples. Furthermore,
recent experiments \cite{Mohiuddin09} did not find any significant
dependence of $\mu$ on immersing graphene devices in high-$\kappa$
media such as ethanol and water (dielectric constants $\kappa
\approx 25$ and $80$, respectively) but this also disagrees with
another report\cite{Fuhrer_PRL08} in which two monolayers of ice
increased $\mu$ in graphene by $\approx 30 \%$. Because of the
experimental controversy, alternative mechanisms such as scattering
on frozen ripples \cite{KG08} and resonant impurities
\cite{KN07,SPG07} were discussed.

Regardless of the experimental debate about the dominant scattering
mechanism, the case of graphene covered with adsorbates at elevated
temperatures \cite{CJAFWI08} generally requires more careful
consideration since there is a vast literature which shows the
formation of clusters of different metals on the surface of
graphite\cite{DM96,BBDP99,YS02,CF05,HB06,CNC08}. These atoms  easily
diffuse on graphite's surface overcoming only relatively low
barriers, and tend to form clusters. Potassium atoms on graphite
arrange themselves into the so called p(2 $\times$ 2) structure,
with a K-K spacing of 0.492 nm, that is, roughly, 3.5
nearest-neighbor carbon-carbon distances \cite{CF05}.
However, in the case of graphite, this usually happens only at low
temperatures and high coverage by adsorbates \cite{CF05}. For low doping
concentrations such as those used in typical experiments on
graphene, adsorbates on graphite are randomly dispersed and, at
elevated temperatures, evaporate from its surface, except for such
materials as, for example, Au, that forms stable clusters on
graphite.

From this surface science perspective, graphene is different from
graphite, and we expect that clusters can be more easily formed on
graphene and be stable at high temperatures. Indeed, it was shown
experimentally \cite{Schedin07} that graphene binds such molecules as NO$_2$,
NH$_2$, H$_2$O, etc. even at room temperature. In the case of
graphite, they can attach only below liquid-nitrogen temperatures\cite{CF05}.
The reason for the stronger attachment remains unclear but could be
due to the presence of ripples on graphene\cite{Metal07}. According to both
experiments and theory \cite{Eetal09}, ripples can bind even atomic hydrogen that
is unstable on a flat surface on both graphene and graphite.

We believe that, once attached to graphene (and this certainly happens for
various gases even at room temperature), adsorbates should tend to cluster,
much more so than for the case of graphite's surface. First, ripples would obviously
force them to move from the valleys onto the hills which favor the adsorption.
Second, there exists an additional long range attraction due to Casimir-like
interaction mediated by Dirac fermions \cite{SAL08}, which is absent for graphite.

On the basis of the above consideration that agrees with what is now known
about graphene adsorbates, both theoretically and experimentally, it is important
to consider how such clustering of adsorbates can
influence the electronic properties of graphene. In this report, we
analyze the scattering of Dirac fermions by clusters of charged
impurities and show that for the same doping level such a disorder
results in significantly lower resistivity. This model reconciles
the doping experiments at cryogenic\cite{CJAFWI08} and
ambient\cite{Schedin07} conditions, as low temperatures prevent
surface diffusion and, therefore, clustering of adsorbates.

The next section presents the model to be studied. Section III
contains the main results. We discuss in section IV possible
extensions of the model. The main conclusions are described in
Section V.

\section{The model} Let us first assume that the charged impurities
inside the cluster are ordered occupying positions over the centers
of carbon hexagons, as in the p(2 $\times$ 2) structure mentioned
above \cite{CF05}. In such a situation the impurities do not break
the sublattice symmetry and cannot lead therefore to the gap
opening. The main effect is therefore merely a local doping of
graphene, that is, shift of its chemical potential, similar to what
happens for graphene on the top of metals \cite{Getal08}. Another
effect, that is, the residual unscreened Coulomb potential, of the
cluster as a whole, $\sim 1/r$, far from the cluster, will be
discussed further.

We start with the simplest model, that is, the scattering of the
charge carriers by a closed region where the chemical potential has
been modified. For simplicity, we assume a circular cluster. The
problem of scattering of the 2D massless Dirac electrons by the
circularly symmetric potential well has been considered in
Refs.\onlinecite{OGM06,HG07,N07,KN07,CPP07,G08}. The model
parameters are the Fermi energy and Fermi wavevector outside the
cluster, $\epsilon_F$ and $k_F$, the change in chemical potential
inside the cluster, $V$, the Fermi velocity, $v_F$, and the radius
of the cluster, $R$. We take $\hbar = 1$ in the following. The
differential cross section can be written in terms of Bessel
functions, whose dimensionless arguments are $\phi_{out} = k_F R$
and $\phi_{in} = ( k_F + V / v_F ) R$. We assume that the cluster is
heavily doped, so that $\phi_{in} \gg \phi_{out}$. The charge
induced inside the cluster is estimated as $\pi ( V R )^2 v_F^{-2}
\propto \phi_{in}^2$. We will neglect intervalley scattering, which
is justified if the boundaries of the cluster are smooth on the
atomic scale, and $R \gg a$, where $a$ is the lattice constant.

\section{Results.} The scattering cross section
reads~\cite{OGM06,HG07,N07,KN07,CPP07,G08}:
\begin{align}
\sigma ( \theta ) &= \frac{4}{\pi k_F} \left| f ( \theta ) \right|^2
\, \, \, \, \, \, \, \, \, \,
 f ( \theta ) =
\sum_{n=-\infty}^{n=\infty} \frac{R_n e^{i n \theta}}{i + R_n}
\nonumber \\R_n &= - \frac{J_n ( \phi_{out} ) J_{n+1} ( \phi_{in} )
- J_{n+1} ( \phi_{out} ) J_n \left( \phi_{in} \right)}{Y_n (
\phi_{out} ) J_{n+1} \left( \phi_{in} \right) - Y_{n+1} ( \phi_{out}
) J_n \left( \phi_{in} \right)} \label{amplitudes}
\end{align}

\begin{figure}
\begin{center}
\includegraphics*[width=6cm,angle=0]{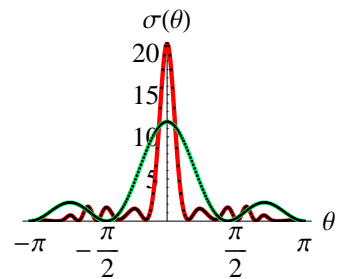}
\caption{Angular dependence of the cross section, $\sigma ( \theta
)$, in nanometers, for a cluster of radius $R=20$ nm with a chemical
potential of $V=500$ meV.  Red: charge density $\rho = 5 \times
10^{12}$cm$^{-2} ( E_F = 250$meV, $k_F R = 7.9 )$. Green: Angular
dependence of the cross section (multiplied by 100) for $\rho =
10^{10}$cm$^{-2} ( E_F = 11$meV, $k_F R = 0.35 )$.}
\label{angular}\end{center}
\end{figure}

Note that, since $R_n = R_{-1-n}$, the back-scattering amplitude
vanishes, $f( \theta = \pi ) = 0$ which is the consequence of the
pseudospin conservation at the ``chiral'' scattering related with
the Klein paradox \cite{KNG06}.

The cross section shows two regimes, depending on whether
$\phi_{out} = k_F R \ll 1$ or $\phi_{out} \gg 1$. In the first case,
the cluster is small compared to the Fermi wavelength. The cluster
perturbs weakly the electronic wavefunctions, and the Born
approximation can be used. The differential cross section, $\sigma (
\theta )$ has in this case a weak dependence on the scattering angle
$\theta$. The total cross section increases as $k_F$ is increases,
$\sigma \sim [ V / ( v_F R^{-1} ) ]^2 k_F R^2$.

For $k_F R \gg 1$, the cross section as function of the incident
angle $\theta$ shows a narrow maximum at $\theta = 0$. In
addition, both the angular resolved and the integrated cross
sections show resonances, associated to quasi-bound states inside
the cluster. The integrated cross section decays slowly as a
function of $k_F$. The angular dependence of the cross section is
shown in Fig.~\ref{angular}.

\begin{figure}
\begin{center}
\includegraphics*[width=6cm,angle=0]{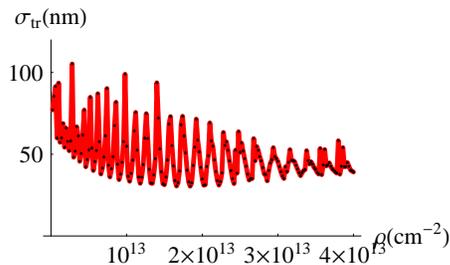}
\caption{Integrated transport cross section $\sigma_{tr}$,  for a
cluster of radius $R=20$ nm and a  shift in the chemical potential
of $V=0.5$ eV.  } \label{total}\end{center}
\end{figure}

Results for the transport cross section, $\sigma_{tr} =
\int_{-\pi}^{\pi} \sigma ( \theta ) [ 1 - \cos ( \theta ) ] d
\theta$, are shown in Fig.~\ref{total}. We analyze in
Fig.~\ref{total} the total cross section for t $V = 0.5$ eV, which
describes the shift in chemical potential due to weakly coupled
adsorbates, like Al, Ag, or Cu \cite{Getal08}. Similar results,
although with a smaller periodicity, are found for $V=2$ eV, which
describes strongly coupled adsorbates, such as $K$, where the charge
transfer can reach 1/8 per carbon atom \cite{DD81}. The radius of
the cluster was chosen as $R=20$ nm, which is comparable to the size in
ripples found in graphene\cite{Metal07}. The total number of electrons
inside the cluster is therefore $N_{in} = \pi \rho R^2 \approx 250$,
where $\rho$ is the charge density inside the cluster, $\rho =
{k_F^{cl}}^2 / \pi , v_F k_F^{cl} = V$.

The limit $k_F R \gg 1$ can be analyzed by using the asymptotic
expressions for the Bessel functions at $x \rightarrow \infty$:
\begin{align}
J_n ( x ) + i Y_n ( x ) &\approx \sqrt{\frac{2}{\pi x}} e^{i \left(
x - \frac{n \pi}{2} - \frac{\pi}{4} \right)}
\end{align}
Then, the expression for the reflection coefficient in radial
waves, $r_n$ (see Ref. \onlinecite{G08}) simplifies to
\begin{equation}
r_n \approx \left\{ \begin{array}{lr} \tan ( \phi_{in} - \phi_{out}
) = \tan \left( \frac{V R}{v_F} \right) &n \ll \phi_{in} \\ 0 &n \gg
\phi_{in}
\end{array} \right.
\end{equation}
and the cross section can be approximated as
\begin{equation}
\sigma ( \theta ) \approx \sum_{n = -n_{max}}^{n=n_{max}} \sum_{n' =
-n_{max}}^{n'=n_{max}} \frac{4 \left| \sin \left( \frac{V R}{v_F}
\right) \right|^2}{\pi k_F} e^{i ( n - n' ) \theta}
\end{equation}
where $n_{max} \sim k_F R$. The transport cross section in this
regime is
\begin{equation}
\sigma_{tr} = \int_{-\pi}^{\pi} \sigma ( \theta ) \left[ 1 - \cos
( \theta ) \right] d \theta \propto \frac{\left| \sin \left(
\frac{V R}{v_F} \right) \right|^2}{k_F} \label{sigma_app}
\end{equation}

\begin{figure}
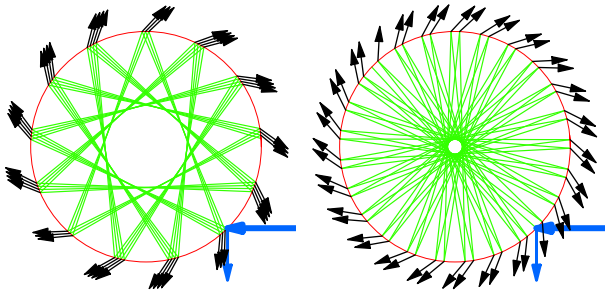

\begin{center}
\includegraphics*[width=4cm,angle=0]{trajectories_cluster_v_1.dat}
\includegraphics*[width=4cm,angle=0]{trajectories_cluster_v_10.dat}
\caption{Classical trajectories of an electron scattered by a
circular cluster. 50 internal reflections are shown. The impact
angle, $\theta$, of the incoming trajectories is $\theta = \pi / 4$.
Left: $V=E_F$. Right: $V=10 E_F$.} \label{trajectories}\end{center}
\end{figure}

\begin{figure}
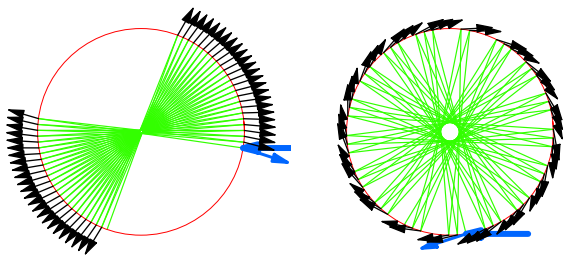

\begin{center}
\includegraphics*[width=4cm,angle=0]{trajectories_cluster_v_10_05.dat}
\includegraphics*[width=4cm,angle=0]{trajectories_cluster_v_10_45.dat}
\caption{As in Fig.~\ref{trajectories}, for $V=10 E_F$, as function
of the impact angle.  Left: $\theta = \pi / 20$. Right: $\theta =
\pi/2 - \pi/20$.} \label{trajectories_angle}\end{center}
\end{figure}

The scattering process in this limit can be studied by the methods
of geometrical optics~\cite{CPP07,CFA07,GCN08}. Typical
trajectories, as a function of the shift in potential inside the
cluster and impact angle are shown in
Figs.~\ref{trajectories}.
The scattering will be dominated by periodic orbits inside the
cluster. These periodic orbits are the semiclassical analogues of
the resonances of the quantum model. For energies such that the
internal trajectories are not periodic, the transmitted waves will
interfere destructively. A periodic trajectory will lead to
transmitted rays at well defined angles, as found in the full
calculation of $\sigma ( \theta )$. Typical trajectories, as
function of the shift in potential inside the cluster and impact
angle are shown in Figs.~\ref{trajectories}
and~\ref{trajectories_angle}. The only periodic orbits for large
values of $V/E_F$ include many internal reflections, which
correspond to high angular momenta in the quantum model. These
orbits are probably less efficient in modifying the scattering
process than the orbits with a lower number of internal reflections,
leading to the calculated cross section, with a sharp maximum as function of
the incident angle. Note that the resonances under discussion are
two-dimensional analogs of the ``Fabry-Perot'' resonances in the
Klein tunneling regime \cite{KNG06}.

The elastic electron mean free path, $l$, is given, approximately
by
\begin{equation}
l \sim \frac{1}{n_C \sigma_{tr}}
\end{equation}
where $n_C$ is the cluster concentration. At low carrier
densities, $k_F R \ll 1$ the Born approximation gives:
\begin{equation}
 \sigma_{tr} \propto k_F R^2 \left( \frac{V}{v_F R^{-1}} \right)^2
\end{equation}
and $\sigma_{tr}$ is proportional to the density of states and to the square of
the potential.
 At high densities, $k_F R \gg 1$, one can use
Eq.(\ref{sigma_app}). The conductivity is estimated as
\begin{equation}
g = \frac{e^2}{h} k_F l \sim \left\{ \begin{array}{lr} \frac{e^2}{h}
\frac{1}{n_C R^2} \left( \frac{v_F R^{-1}}{V} \right)^2 & k_F R \ll 1 \\
\frac{e^2}{h} \frac{k_F^2 }{n_C} & k_F R \gg 1 \end{array} \right.
\label{conductivity}
\end{equation}
We expect the oscillations of the cross section shown in
Fig.~\ref{total} to be averaged out in clusters with less symmetric
shapes. The parameter $k_F R$ reaches the value $k_F R \approx
10-12$ for $R=20$ nm and charge density in the clean regions $\rho =
2 \times 10^{13}$cm$^{-2}$.

Interestingly, for the regime $\phi_{out} \gg 1$ the total cross
section $\sigma_{tot}$ distinguished from $\sigma_{tr}$  by the
absence of the factor $1 - \cos{\theta}$ in Eq. (\ref{sigma_app})
is larger than $\sigma_{tr}$ by a factor $k_F R$. The total cross
section is related with the single-particle decoherence time which
determines, e.g., Dingle temperature in the Shubnikov - de Haas
oscillations \cite{dingle}.

The elastic mean free path depends on the cluster density and
carrier concentration, $\rho$. For $n_C = 10^{10}$cm$^{-2}$ and
$\rho = 5 \times 10^{12}$ cm$^{-2}$ we obtain $l = 1/(n_C \sigma)
\approx 200$ nm.

We have neglected so far the long-range part of the Coulomb
potential induced by the cluster. This potential will modify the
scattering cross section for electron wavelengths $k_F^{-1}
\gtrsim R$. The cross section will depend on carrier concentration
as $\sigma_{tr} \propto k_F^{-1} $ \cite{N07,SKL07,FNS07,PNN07}.
As a result, we expect that the conductivity for $k_F R \ll 1$
will scale as $k^{2}_{F}$ \cite{NM07,AHGS07}, instead of the
dependence given by Eq.(\ref{conductivity}). However, since the
scattering cross section is proportional, in Born approximation,
to the charge square and to the first power of the charge
concentration, the clusterization will lead to suppression of this
contribution to the resistivity by a factor of order of number of
atoms in cluster, in comparison with the case of chaotically
distributed impurities.

\begin{figure}
\begin{center}
\includegraphics*[width=6cm,angle=0]{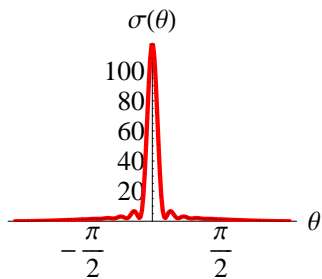}
\caption{Angular dependence of the cross section when the cluster is
determined by a mass term, which breaks the symmetry between the
sublattices. The parameters used are $k R = 10$ and $\Delta R / v_F
= 20$.} \label{mass}\end{center}
\end{figure}

\section{ Beyond the simplified model} Our model of completely
ordered impurities inside the cluster is oversimplified. However, if
disorder inside the cluster is relatively weak so that the {\it
local} mean free path $l$ exceeds the electron wavelength inside the
cluster $\lambda \approx h v_F /V$ one can expect that above
consideration is correct, at least, qualitatively (the local mean
free path is defined here as the mean free path of infinite
disordered system with the same chemical potential and the same
distribution of the scattering potential as inside the cluster). If
the disorder becomes stronger one reaches at some moment the Mott
limit $l \approx \lambda$ without further localization, due to the
Klein tunneling \cite{KNG06}. In this regime, the electron rays
inside the cluster are no more straight and the ``Fabry-Perot''
resonances are destroyed. The cluster with such strong disorder will
behave just as an obstacle of size $R$, with the transport cross
section of order of $R$.

Another effect which should be considered is a possible formation
of superstructure inside the cluster. It can break the sublattice
equivalence and lead to the local gap opening. To see potential
consequences of this local reconstruction of the electronic
structure one can extend the model to the case when the cluster is
defined by a mass term rather than a shift of chemical potential.
Similar boundary conditions were discussed in Ref.
\onlinecite{BM87}. We assume that the mass, $\Delta$, is only
finite inside the cluster. We also neglect, for simplicity, by the
shift of the chemical potential. The cross section in such model
is expressed in terms of the new reflection amplitudes (cf.
Eq.(\ref{amplitudes})):
\begin{align}
 R_n &= \nonumber \\ &- \frac{ i a_- J_n ( k_F R ) I_{n+1} ( \kappa
R ) - a_+ J_{n+1} ( k_F R ) I_n ( \kappa R )}{i a_- Y_n ( k_F R )
I_{n+1} ( \kappa R ) - a_+ Y_{n+1} ( k_F R ) I_n ( \kappa R )} \nonumber \\
\kappa &=\frac{\sqrt{\Delta^2 - (E_F + V)^2}}{v_F} 
\, \, \, \, \, \, \, \, \,
a_\pm = \sqrt{\frac{1}{2} \pm
\frac{\Delta}{2 E}} \label{amplitudes_mass}
\end{align}
where $I_n ( x )$ is a modified Bessel function, which is zero at
the origin and grows exponentially as $x \rightarrow \infty$.

We have also calculated the cross section including  a staggered
potential, $\Delta$. The main effect of a mass term seems to be to
reduce the oscillations of the transport cross section as a function
of angle. If the mass term is large enough, the effect should be
qualitatively the same as for the strong disorder, that is, the
transport cross section will be of the order of $R$, as for a
nontransparent obstacle in optics. The changes induced by a mass
term in the differential cross section are shown in Fig.~\ref{mass}.

\section{Conclusions.} Let us summarize the main results of our
consideration.

(i) The transport cross section of charge carriers in graphene by
large neutral clusters due to a shift of the chemical potential
inside the cluster becomes independent of the cluster size, $R$, and
shift in chemical potential, $V$, for $k_F R \gg 1$, except for an
oscillatory function. This can be viewed as a consequence of the
Klein tunneling \cite{KNG06}: electrons can always tunnel into the
cluster, irrespective of the value of $V$. The oscillatory function
is, most likely, replaced by its average for clusters with irregular
shapes, as one can assume by analogy with the geometric optics
\cite{DS93}.

(ii) The total scattering cross section, obtained by integrating
$\sigma ( \theta )$ over angles, is proportional to $R$ for $k_F R
\gg 1$, as it should. In this regime $\sigma_{tot} / \sigma_{tr} \approx k_F R \gg
1$ which, is principle, can be observed by comparison of the
mean free path with the Dingle temperature if this scattering
mechanism is dominant. For all other scattering mechanisms
considered before, including charged impurities, $\sigma_{tot}
\approx \sigma_{tr}$, with a numerical factor of order of one.

(iii) The transport cross section is proportional to $k_F^{-1}$.
Hence, scattering by large clusters leads to a dependence on
carrier density similar to that for charged impurities or resonant
scatterers, $g \propto n$.

(iv) The main difference in the expression for the conductivity
between scattering by neutral clusters and scattering by charged
impurities is that the impurity concentration has to be replaced by
the cluster concentration which increases the electron mobility,
roughly, by two orders of magnitude. Thus, possible clusterization
of charged impurities in graphene can probably explain the relatively weak
dependence of the mobility on charge impurity concentration
\cite{Schedin07} and dielectric constant\cite{Mohiuddin09}.

v) The formation of clusters is a process favored by high atomic
diffusion. Hence, we expect that, by
annealing the samples used in \cite{CJAFWI08} above 100K
the mobility will increase towards the values measured before doping by potassium.

vi) The correlation observed in\cite{Petal09} between the shift of
the Dirac point and the electron mobility for different adsorbates
as a function of adsorbate concentration is consistent with the formation of
clusters. The effective charge, $q^*_i$, transferred from the adsorbate atom
to the graphene layer varies for different adsorbates. For elements that transfer
to graphene an amount of charge much less than one electron charge,
such as Pt, the scattering cross section\cite{N07} goes as ${q^*_i}^2$. The
shift of the Dirac point should be proportional to ${E_D}_i \propto n_i q_i^*$, where $n_i$
is the concentration of the adsorbate. The change in mobility should scale, on the other hand, as $\mu_i^{-1}
\propto n_i {q_i^*}^2$. For adsorbates such that $q_i^* \approx 1 e$, the mobility
scales as $\mu_i^{-1} \propto n_i {q_i^*}$. Hence, different adsorbates should show
different ratios  ${ \mu^{-1}_i / E_D}_i$. A ratio that varies weakly
for different adsorbates is more consistent with the existence of clusters, each
of which transfers to graphene a few free electron charges, independently of
the type of adsorbate and size of the cluster.

\section{Acknowledgements} We acknowledge interesting conversations with R. K. Kawakami
and S.-T. Tsai. FG acknowledges support from MEC (Spain) through
grant FIS2005-05478-C02-01 and CONSOLIDER CSD2007-00010, and by the
Comunidad de Madrid, through CITECNOMIK, CM2006-S-0505-ESP-0337.
This research was supported in part by the National Science
Foundation under Grant No. PHY05-51164. MK acknowledges support from
Stichting voor Fundamenteel Onderzoek der Materie (FOM), the
Netherlands.

\bibliography{bib_scattering_2}
\end{document}